\documentclass[conference, 9pt]{IEEEtran}
\usepackage{amsmath,amssymb,amsfonts}
\usepackage{graphicx}
\usepackage{textcomp}
\usepackage{marvosym} 
\usepackage{ragged2e}
\usepackage{algpseudocode}
\usepackage{tabularx}
\usepackage{tcolorbox}
\usepackage{caption}
\usepackage {subcaption}

\tcbuselibrary{skins}

\usepackage{multirow}
\usepackage{makecell}
\usepackage{booktabs}
\usepackage{threeparttable}
\usepackage{diagbox}

\usepackage{CJKutf8}

\usepackage{stfloats}

\def\BibTeX{{\rm B\kern-.05em{\sc i\kern-.025em b}\kern-.08em
    T\kern-.1667em\lower.7ex\hbox{E}\kern-.125emX}}

\def\eg{\textit{e.g.,}\ }
\def\ie{\textit{i.e.,}\ }

\newcommand{\cmd}[1]{\texttt{\small{#1}}}

\usepackage{algorithm}
\usepackage{algorithmicx}
\usepackage{algpseudocode} 

\usepackage{multicol}
\usepackage{listings}
\usepackage{tikz}
\usepackage{caption}
\usepackage{subcaption}

\usepackage{caption}
\usepackage{listings} 
\usepackage{multirow}

\DeclareCaptionFormat{custom}{#3}
\usepackage{xcolor}
\usepackage{tabularx}
\usepackage{tcolorbox}
\usepackage{caption}
\usepackage {subcaption}

\tcbuselibrary{skins}

\lstdefinelanguage{json}{
    basicstyle=\normalfont\ttfamily,
    numbers=left,
    numberstyle=\scriptsize,
    stepnumber=1,
    numbersep=8pt,
    showstringspaces=false,
    breaklines=true,
    frame=lines,
    backgroundcolor=\color{white},
    literate=
     *{0}{{{\color{blue}0}}}{1}
      {1}{{{\color{blue}1}}}{1}
      {2}{{{\color{blue}2}}}{1}
      {3}{{{\color{blue}3}}}{1}
      {4}{{{\color{blue}4}}}{1}
      {5}{{{\color{blue}5}}}{1}
      {6}{{{\color{blue}6}}}{1}
      {7}{{{\color{blue}7}}}{1}
      {8}{{{\color{blue}8}}}{1}
      {9}{{{\color{blue}9}}}{1}
      {:}{{{\color{purple}:}}}{1}
      {,}{{{\color{purple},}}}{1}
      {\{}{{{\color{brown}\{}}}{1}
      {\}}{{{\color{brown}\}}}}{1}
      {[}{{{\color{brown}[}}}{1}
      {]}{{{\color{brown}]}}}{1},
}

\usepackage{url}

\begin{document}
\begin{CJK*}{UTF8}{gbsn}


\title{Wit-HW: Bug Localization in Hardware Design Code \\ via Witness Test Case Generation}


\author{

\IEEEauthorblockN{Ruiyang Ma\textsuperscript{1}, Daikang Kuang\textsuperscript{1}, Ziqian Liu\textsuperscript{2}, Jiaxi Zhang\textsuperscript{1,\Letter}, Ping Fan\textsuperscript{3}, Guojie Luo\textsuperscript{1,4,5,\Letter}}

\IEEEauthorblockA{\textit{\textsuperscript{1}School of Computer Science, Peking University; \textsuperscript{2}School of Information, Renmin University of China; \textsuperscript{3}Deepoly Technology Inc.}}
\IEEEauthorblockA{\textit{\textsuperscript{4}National Key Laboratory for Multimedia Information Processing, Peking University}}
\IEEEauthorblockA{\textit{\textsuperscript{5}Center for Energy-efficient Computing and Applications, Peking University}}

\IEEEauthorblockA{ruiyang@stu.pku.edu.cn, zhangjiaxi@pku.edu.cn, gluo@pku.edu.cn}

\vspace{-2em}
}







\maketitle
\thispagestyle{empty}

\begin{abstract}

Debugging hardware designs requires significant manual effort during hardware development. After engineers identify a bug-triggering test case in simulation-based hardware verification, they usually spend considerable time analyzing the execution trace to localize the bug.
Although numerous automated hardware debugging techniques exist, they are not applicable to large designs and deep bugs. A primary reason for their limitations is that these techniques only utilize the information of a single bug-triggering test case for bug localization, which prevents them from effectively analyzing intricate hardware systems and figure out the root cause of bugs.
To solve this problem, in this paper, we transform the hardware bug localization problem into a test generation problem, aiming to find a set of effective witness test cases beyond the initial bug-triggering test case to enhance hardware bug localization. Witness test cases refer to the cases that do not trigger the bug in the faulty design. By analyzing the execution differences between passing and failing test cases with spectrum-based method, we can eliminate innocent design statements and localize the buggy ones.
To further refine the suspicious area, we define the criteria for effective witness test cases and use a mutation-based strategy to generate such test cases. 
Based on this approach, we propose an automated hardware bug localization framework named Wit-HW. We evaluate Wit-HW on 41 bugs from various hardware designs. The experimental results show that Wit-HW effectively localize 49\%\,/\,73\%\,/\,88\% bugs within Top-1\,/\,Top-5\,/\,Top-10 ranks, significantly outperforming state-of-the-art bug localization techniques. Additionally, we evaluate Wit-HW on 13 real-world bugs collected from open-source hardware projects, showcasing the robust performance of our method.

\end{abstract}






\section{Introduction}
Hardware bugs in modern circuit systems can lead to significant losses. Finding and resolving these bugs during hardware development process is challenging. Although researchers have devoted dedicated efforts, hardware verification is still a tedious and time-consuming process~\cite{bg_1}.
In simulation-based hardware verification, the RTL model is simulated and its behavior is compared with a golden model~\cite{bg_2}. If there is a difference in behavior, a bug is triggered. Due to the complexity of hardware designs, engineers often spend hours or even days to localize and fix the bug. In hardware development, over 70\% of the time is devoted to verification~\cite{bg_3}, with approximately 50\% of that time spent on debugging~\cite{siemens_study}.

Numerous automatic bug localization techniques have been proposed in the field of software debugging, such as slicing-based techniques and spectrum-based techniques (SBFL)~\cite{sw_bug_loc_survey}. However, these techniques encounter quite a few difficulties when applied to hardware designs. 
Firstly, unlike the serial execution in software, hardware designs feature concurrent structures and strong interconnections between signals. This complexity results in a much larger suspicious buggy scope when using slicing-based methods~\cite{cirfix}. 
Secondly, hardware designs typically operate as sequential models. Errors are always triggered (observed in output signals) out of sync with the execution of buggy statements (recorded by coverage report). This misalignment complicates the process when identifying the root cause of the bug with spectrum-based methods~\cite{tarsel}. 

In previous research on automatic hardware debugging, CirFix~\cite{cirfix} and RTL-Repair~\cite{rtl_repair} use slicing-based method to localize bug and automatically modify the suspicious code segments with repair templates. They aim to find repair solutions that ensure the design behaves correctly under the given test case. However, due to the first difficulty mentioned, these approaches exhibit poor performance in large designs.
On the other hand, Tarsel~\cite{tarsel} employs a time-aware spectrum-based method that segments the test case over time and compares code coverage with and without error occurrence. By analyzing the coverage differences, it calculates the statement suspicious probability with statistical methods. Nevertheless, due to the second difficulty, this method struggles to identify bugs that are hidden in sequential designs.

We observe the common limitation of previous research is that they only rely on a single bug-triggering test case for bug analysis and localization. Considering the difficulties mentioned, it is essential to generate more test cases to thoroughly observe the behavior of intricate hardware designs when attempting to localize bugs, which is helpful for both refining the suspicious code scope and identifying the sequential root causes of hardware bugs. 
While many studies focus on hardware test case generation aimed at improving coverage~\cite{rfuzz, hw_fuzz} or finding bugs~\cite{thehuzz, socfuzzer}, none of them are directly applicable to the task of hardware bug localization.

In this paper, we present Wit-HW (Witnessing Hardware), a framework designed to automatically generate a diverse set of effective witness test cases for localizing hardware bugs. 
The concept of \textit{witness test case} is inspired by software bug localization techniques~\cite{ochiai, diwi}, which refers to the passing test cases (\ie those do not trigger the bug) on buggy designs. 
By using a spectrum-based method that employs probabilistic inference to analyze execution trace differences between passing and failing test cases, Wit-HW is able to eliminate suspicion from innocent statements and accurately identifies the buggy parts.
The main contribution of Wit-HW lies in answering two key questions:
\begin{itemize}
\item What types of witness test cases are effective to enhance hardware bug localization performance?
\item How can we efficiently generate these witness test cases?
\end{itemize}

To address the first question, Wit-HW establishes specific metrics to evaluate the effectiveness of witness test cases, which focuses on two criteria: similarity and diversity. 
First, witness test cases should share a similar hardware execution trace to the bug-triggering test case, which helps eliminate more innocent statements from suspicion. 
Second, there should be diversity among the witness test cases in their hardware execution traces to prevent bias during bug localization.
The core of both criteria involves calculating the distance between two hardware execution traces. To achieve this, we design coverage distance and state distance, which comprehensively express the distance and facilitate the measurement of similarity and diversity.

To address the second question, we employ a mutation-based test generation method that utilizes heuristics to create witness test cases meeting specified criteria. 
This process begins by selecting a seed witness test case from the seed set based on similarity criterion, prioritizing those with high similarity with the target. 
Next, we select mutation positions based on diversity criterion, favoring those that have previously shown greater capacity for generating diversity within the seed set.
Finally, we add the newly generated test case into the seed set, facilitating an iterative process that progressively improves the quality of the witness test cases.

To evaluate the performance of Wit-HW, we establish benchmarks based on prior works~\cite{cirfix, tarsel, rtl_repair}, which consist of 10 designs and 41 bugs, including some real-world designs and bugs. 
The experimental results show that Wit-HW effectively localize 49\%\,/\,73\%\,/\,88\% bugs within Top-1\,/\,Top-5\,/\,Top-10 ranks, which significantly outperforms Tarsel~\cite{tarsel} (22\%\,/\,39\%\,/\,54\%) and RTL-Repair~\cite{rtl_repair} (15\%\,/\,17\%\,/\,39\%), demonstrating the effectiveness of introducing witness test cases for hardware bug localization. We also investigate the contribution of criteria-guided test generation method compared to a random search strategy, showcasing the effectiveness of our criteria and metrics.
Additionally, we evaluate Wit-HW on 13 real-world bugs collected from open-source FPGA projects, successfully localizing 6 bugs within Top-1 or tied Top-1 ranks, showcasing its robust applicability. Finally, we analyze the limitations of Wit-HW and outlines future directions for improvement.

In summary, this paper makes the following contributions:

\begin{itemize}
\item We propose to transform the problem of hardware bug localization to the problem of witness test case generation.
\item We design similarity and diversity criteria for evaluating the effectiveness of witness test cases.
\item We design specific test generation method to produce a diverse set of effective witness test cases based on the criteria.
\item We perform evaluation and show that Wit-HW significantly outperform existing techniques, which demonstrates the effectiveness of using witness test cases for hardware bug localization.
\end{itemize}

The code is available at \url{github.com/magicyang1573/Wit-HW}.




\section{Background}

\subsection{Spectrum-Based Bug Localization}
In the software domain, many powerful and effective localization approaches have been proposed to automate the bug localization process, such as slicing-based method~\cite{slice_1, slice_2} and spectrum-based method~\cite{ochiai, diwi, sbfl_2}. Slicing-based method abstracts a program into a reduced form by removing irrelevant parts related to the bug, which helps limit the code that the developer needs to inspect~\cite{sw_bug_loc_survey}. However, this method is not particularly strong for automatically localizing the root cause of a bug in a complex system.

Spectrum-based fault localization (SBFL) is a more powerful and popular technique for precisely localizing bugs~\cite{sw_spectrum_survey}. This method collects information about program execution (\eg the statement execution trace). By comparing execution traces from passing and failing test cases, SBFL calculates the suspiciousness of each statement, indicating the likelihood that it contains a bug. 
Intuitively, statements executed more in failing cases are more likely to be buggy, while those executed more in passing cases are more likely to be correct. This executable statement hit spectrum is utilized by popular bug localization techniques like Tarantula~\cite{tarantula} and Ochiai~\cite{ochiai}. 
Consider a statement $s$, where $e_f(s)$ and $n_f(s)$ represent the number of failing test cases that execute and do not execute the statement $s$, and $e_p(s)$ denotes the number of passing test cases that execute the statement $s$. The Ochiai coefficient~\cite{ochiai}, a widely used metric in SBFL~\cite{sw_bug_loc_survey}, calculates the suspiciousness $sus(s)$ of statement~$s$ to be:
\begin{equation}
\label{eq:ochiai}
\text{\textit{sus}}(s) = \frac{e_f(s)}{\sqrt{(e_f(s) + n_f(s)) \cdot (e_f(s) + e_p(s))}}
\end{equation}

By statistical calculations, spectrum-based method helps developers identify suspicious areas. In this paper, we adapt the principle of spectrum-based localization techniques to create Wit-HW framework, which aims to pinpoint buggy statements in an RTL design.

\subsection{Automatic Hardware Debugging}
In the hardware verification process, debugging is a critical and time-consuming task. 
During simulation-based verification, a bug is identified if a test case causes the hardware design to behave differently from its golden reference model or violates certain assertions. Recently, there has been a growing focus on automated hardware debugging, which falls into two main categories: hardware code repair~\cite{cirfix, rtl_repair, llm_fix_1, llm_fix_2} and hardware bug localization~\cite{tarsel, veribug}, where our work targets the latter.

\textbf{Hardware Code Repair: }Hardware automatic code repair methods aim to directly fix the hardware design code, typically written in hardware description languages like Verilog. The objective is to modify the design so that it successfully passes the current testbench. 
CirFix~\cite{cirfix} introduces several structural repair templates to fix the buggy code. It uses a slicing-based method to determine the suspicious code scope by tracing back from the erroneous output signals. Within this scope, CirFix applies genetic programming to iteratively search for correct repairs. 
Building on CirFix, RTL-Repair~\cite{rtl_repair} enhances the code-fixing capability by developing variable-included repair templates. It uses SMT-based methods to solve for variables based on the correct behavior of the design within an adaptive time window. 
Some approaches also utilize large language model (LLM) to repair buggy hardware designs. RTLFixer~\cite{llm_fix_1} employs retrieval-augmented generation and ReAct prompting to enable automatic syntax errors fixing for Verilog code with LLM. \cite{llm_fix_2} uses LLM to fix hardware security bugs with human instructions. 

Current automatic repair works can only effectively fix bugs in simple cases. For large hardware designs or complex deep bugs, CirFix~\cite{cirfix} and RTL-Repair~\cite{rtl_repair} fail to provide correct fixes. LLM-based methods~\cite{llm_fix_1, llm_fix_2} also cannot fully understand the root cause of hardware bugs and typically require human input for bug identification and repair instructions before fixing the bug.
These automatic repair techniques do not perform precise bug localization before attempting repairs. As a result, they must modify a large portion of the design space to identify the exact bug position, which significantly limits their effectiveness in large designs with deep bugs. 
Besides, in typical human debugging, once the bug location is identified, it can usually be easily fixed. 
That is why automatic hardware bug localization is so important. It is extremely necessary to precisely localize hardware bugs to both improve hardware automatic repair and speedup human debugging process. 

\textbf{Hardware Bug Localization: }
Bug localization focuses on identifying the location or root cause of logic errors, without attempting to fix them.  
Tarsel~\cite{tarsel} makes a first attempt to automatically localize hardware bugs with a time-aware spectrum-based localization technique. Given a bug-triggering test case, it samples coverage near the error occurrence time, compares coverage differences between error and non-error time, and calculates suspicious statements. However, by analyzing only one test case, Tarsel struggles to capture the root cause of many deep bugs in sequential designs.
VeriBug~\cite{veribug} uses deep learning to calculate the attention weight of each statement, capturing its effect on the target output. It compares the attention differences between correct and buggy simulation traces to identify suspicious statements. Nonetheless, 
obtaining comprehensive circuit training data during normal hardware development is difficult, which limits the practical application of this method.

Most existing automatic bug repair and localization techniques only rely on a single bug-triggering test case for analysis. This approach lacks the ability to thoroughly observe the behaviors of buggy hardware designs, thereby severely limiting their performance. 
In this paper, we introduce Wit-HW, which offers a novel perspective that transforms this paradigm. We focus on generating effective witness test cases to provide sufficient evidence for comprehensive bug analysis, which will be detailed in the following sections.






\section{Problem Formulation}\label{sec:formulation}
Wit-HW transforms hardware bug localization based on a single test case into the problem of hardware test case generation. We model this as witness test case generation problem under effectiveness criteria. These generated test cases enable spectrum-based bug localization techniques to achieve higher accuracy compared to relying solely on a single bug-triggering test case or randomly generated test cases.
In this section, we introduce the problem of hardware bug localization and its translation to witness test case generation.  

\vspace{-2pt}
\subsection{Hardware Bug Localization}


Suppose the buggy RTL design $b$ is composed of \( z \) statements \( S = \{s_1, s_2, \ldots, s_z\} \), in which some statements \( S_b \subseteq S \) have bugs. The design has \( n \) input signals. A test case \( C \) given to the design contains the values of \( n \) input signals over \( t \) clock cycles. Given the test case \( C \), the buggy design produces an output \( O_b(C) \).
We also have an output validator $V$, which is used to determine whether \(O_b(C)\) meets the expected behavior.  If \(O_b(C)\) is deemed incorrect by the validator, a bug is triggered, indicating that some statements in the RTL design are faulty.

The goal of bug localization is to identify which statement is buggy. In spectrum-based methods, the output is a suspiciousness score list \( R = \{r_1, r_2, \ldots, r_z\} \), which reflects the likelihood of each statement containing a bug. 
Now the problem formulation of hardware bug localization can be summarized as follows:

\textit{Input: }
Buggy design $b$, validator $V$, and one failing test case \( C_b \).

\textit{Output: }
Statement suspiciousness scores \( R \).


\textit{Objective:} Establish an ordering $R$ s.t. $\forall s_i{\in}S_b, \forall s_j{\notin}S_b$, $r_i > r_j$.



\vspace{-2pt}
\subsection{Witness Test Case Generation}
Traditional SBFL localizes bugs by analyzing trace differences between failing and passing test cases. Relying solely on the bug-triggering test case is imprecise.
Passing test cases can serve as witness to eliminate innocent parts from executed statements, which are also called witness test cases. 
Therefore, it is useful to decompose hardware bug localization as two stages: 1) witness test case generation, and 2) bug localization with witness test cases.

\textbf{Stage-1:} The goal of this stage is to generate a set of witness test cases \( \mathcal{P} = \{C_1, C_2, \ldots, C_m\} \) from the failing one.
However, not all witness test cases are effective for SBFL. 
For example, generating an illegal test case make the faulty design exhibit the same behavior as the golden model, but such witness test case cannot reflect meaningful hardware behavior and is useless for SBFL. Therefore, it is important to define appropriate effectiveness metrics to assess witness test cases. Wit-HW develops metrics based on similarity and diversity criteria (detailed in Section~\ref{sec:overview}).
With these measurable metrics, the task can be framed as an optimization problem aimed at generating witness test cases that best meet the effectiveness metrics.

\textit{Input: }
Buggy design $b$, validator $V$, and one failing test case \( C_b \).

\textit{Output: }
A set of effective witness test cases \(\mathcal{P}\).


\textit{Objective: }
Find \( \mathcal{P} = \{ C_i \} \) to maximize \(\text{\textit{eff. metrics}}(\mathcal{P})\).

\textbf{Stage-2:} With the set of effective witness test cases \(\mathcal{P}\), we can extract the execution traces from these cases and apply SBFL methods to compute more accurate statement suspiciousness scores \( R \), which translates the problem back to hardware bug localization.

\textit{Input: }
Execution traces of witness cases \(\mathcal{P}\) and failing case \( C_b \).

\textit{Output: }
Statement suspiciousness scores \( R \).

In the two-stage process of Wit-HW, the key is the effectiveness criteria of witness test cases, which directly impact the generation of witness test cases in the first stage and the bug localization accuracy of SBFL in the second stage. 
In the following sections, we will introduce our designed criteria and demonstrate its effectiveness.


\section{Witness Test Case Criteria and Generation Method}
\label{sec:method}

\subsection{Overview of Wit-HW}
\label{sec:overview}
\begin{figure}[t] 
    \centering 
    \includegraphics[width=0.45\textwidth]{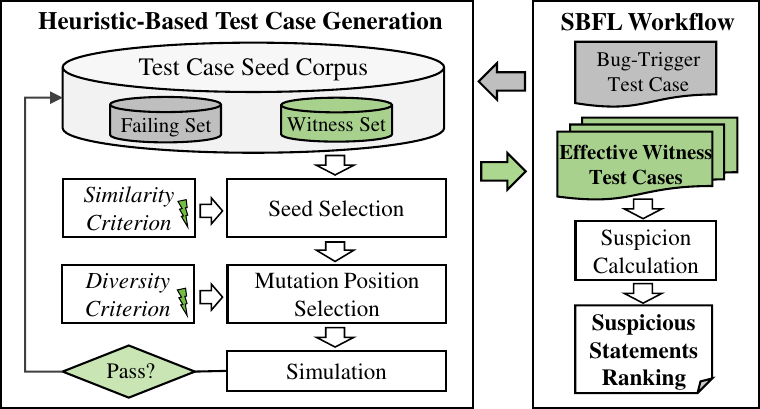} 
    \caption{Overview of Wit-HW.} 
    \label{fig:overview} 
    \vspace{-15pt}
\end{figure}

We introduce the framework Wit-HW for localizing hardware bugs. The overview of Wit-HW is presented in Figure~\ref{fig:overview}.
Wit-HW involves automatically generating a set of effective \emph{witness test cases} (\ie test cases that do not trigger the bug), and then comparing their execution traces with the given bug-triggering test case using SBFL to localize bugs.
Two key challenges are critical to the success of Wit-HW.
First, we must carefully construct witness test cases to avoid introducing low-quality test cases, which can create excessive noise. 
Second, the witness test cases should be distinct enough to eliminate innocent suspects, otherwise the suspiciousness calculation may become biased. 
To address these challenges, we define two criteria for constructing an effective set of witness test cases:


\begin{itemize}
    \item \textbf{Similarity Criterion:} Each test case in the set should share a similar hardware execution trace with the failing test case.
    \item \textbf{Diversity Criterion:} The test cases in the set should exhibit diversity in their hardware execution traces.
\end{itemize}

However, given the extensive space of potential witness test cases, generating an effective set that meets the two key criteria is challenging. 
Wit-HW designs measurable metrics based on these criteria and employs these metrics as heuristics in the evolutionary mutation-based test case generation process.
To satisfy the first criterion, we develop distance metrics that accurately and efficiently measure the similarity between the hardware execution traces of witness test cases and bug-triggering test cases, which is detailed in Section~\ref{sec:similarity}.
To satisfy the second criterion, we introduce priority metrics to reflect the diversity effect of mutation position selection, as detailed in Section~\ref{sec:mutate}.
Subsequently, we incorporate these metrics as heuristics to guide the mutation-based witness test case generation, which is explained in Section~\ref{sec:heuristic}.
Finally, we calculate the suspiciousness of each statement using the effective witness test cases and the bug-triggering test case, which is detailed in Section~\ref{sec:sbfl}.

\subsection{Similarity Criterion: Distance Metrics for Hardware Execution}
\label{sec:similarity}
An effective witness test case should exhibit similar hardware execution behavior to the bug-triggering test case while not actually triggering the bug. This allows SBFL to compare the statement execution status of witness test cases with bug-triggering test cases and more accurately identify suspicious statements. We define two distance metrics that comprehensively measure the similarity between the hardware execution traces of two test cases.

\subsubsection{Coverage Distance} 
Coverage indicates the hit count of each statement during the simulation of a test case. Suppose we have two test cases: the witness test case \( C_w \) and the bug-triggering test case \( C_b \) written by the verification engineer. Their respective coverage reports are \( H_w = \{ h_{w}^1, h_{w}^2, \ldots, h_{w}^z \} \) and \( H_b = \{ h_b^1, h_b^2, \ldots, h_b^z \} \), where \(z\) is the line number of statements in the hardware design. We define the coverage distance as:
\begin{equation}
\label{eq:cov_dis}
D_{cov}(C_w, C_b) = \sqrt{\sum\nolimits_{i=1}^{z} (\, h_w^i - h_b^i \,)^2}
\end{equation}
Using coverage distance as metric ensures that the generated witness test cases touch similar code parts as the bug-triggering test case, thereby eliminating as many innocent statements from suspicion as possible. However, hardware designs are typically sequential models in terms of timing dependencies, similar coverage reports do not necessarily imply similar execution behavior; the hit orders of statements may differ. Therefore, additional metrics are needed to provide a more comprehensive evaluation.

\subsubsection{State Distance} 
To reflect the sequential nature of hardware execution, a straightforward approach is to compare the concrete hit status of each statement across clock cycles. However, this information cannot be directly acquired by hardware simulators and calculating such metric can be computationally expensive for large designs with many statements and long input sequences. To make the metric both feasible and reflective of hardware execution behavior, we focus on the value transitions of the registers.

The value transition sequence can be represented as \( Q = \{ (t_1, v_1), \\ (t_2, v_2), \ldots \} \). Each tuple indicates the variable transits value to \(v\) at time \(t\). We employ Dynamic Time Warping (DTW)~\cite{dtw} to measure the similarity between two state transition sequences. Unlike traditional distance metrics, DTW aligns sequences in a non-linear manner, allowing for flexible matching of values that may occur at different times. 
Given two value transition sequences \( Q_w \) and \( Q_b \) from test cases \( C_w \) and \( C_b \), DTW algorithm computes a cost matrix \( D \), where each element \( D(i, j) \) represents the cumulative cost of aligning the first \( i \) elements of \( Q_w \) with the first \( j \) elements of \( Q_b \). The recurrence relation used to fill the cost matrix is defined as:
\begin{equation}
D(i, j) = (\,Q_w[i] \neq Q_b[j]\,) + \min \begin{cases}
D(i{-}1, j) \\
D(i, j{-}1) \\
D(i{-}1, j{-}1)
\end{cases}
\end{equation}
The lengths of sequences \( Q_w \) and \( Q_b \) are denoted by \( l_w \) and \( l_b \). The final state distance between the two state transition sequences can be obtained from the last cell of the cost matrix:
\begin{equation}
DTW(Q_w, Q_b) = D(l_w, l_b)
\end{equation}
For the buggy design with \( k \) registers, where \( Q_w^i \) and \( Q_b^i \) reflect the value transitions of register \( i \), we define state distance metric as:
\begin{equation}
D_{state}(C_w, C_b) = \sqrt{\sum\nolimits_{i=1}^{k} (DTW(Q_w^i, Q_b^i))^2}
\end{equation}

Finally, we can express the overall distance between the witness test case \( C_w \) and the bug-triggering test case \( C_b \) using a weighted combination of the coverage distance and the state distance metrics:
\begin{equation}
\label{eq:distance}
Distance(C_w, C_b) = \alpha \cdot D_{cov}(C_w, C_b) + \beta \cdot D_{state}(C_w, C_b)
\end{equation}

\subsection{Diversity Criterion: Priority Metrics for Mutation Position}
\label{sec:mutate}

In Wit-HW, our objective is to generate effective witness test cases based on the initial bug-triggering test case. We have defined the hardware execution distance metric between two test cases in Section~\ref{sec:similarity}. In this subsection, we use this metric to assess diversity and focus on generating diverse witness test cases through mutation.



Based on a selected seed test case with $t$ clock cycles, Wit-HW selects several time positions and performs mutation operations on these positions (detailed in Section~\ref{sec:heuristic}). However, not all mutation positions are equally effective in generating diverse witness test cases. Therefore, we design an adaptive procedure to prioritize mutation positions that contribute to the diversity of witness test cases.

Intuitively, if a mutation position frequently generates witness test cases with greater diversity compared to existing ones, such mutation position should be selected with higher probability for further mutations. Based on the intuition, we maintain a priority for each mutation position. Initially, each position has the same priority \(\delta\). We calculate the score of a generated test case \( C'\) as:
\begin{equation}
    score(C') = \frac{1}{m}\sum\nolimits_{i=1}^{m}Distance(C', C_i)
\end{equation}
\(C_i\) refers to one of the \(m\) seed witness test cases currently stored in the seed set (detailed in Section~\ref{sec:heuristic}). We assess the similarity of \(C'\) with those in the seed set using the distance metric. A high score indicates \(C'\) is different from existing ones, contributing to diversity. 

After getting the score of a mutation, we update the priority for positions mutated in \(C'\). Since the selection of mutation positions depends on recent status of the seed set, we update the priority score with Exponential Weighted Moving Average (EWMA)~\cite{ewma}. The priority for mutation position $i$ is updated by the following equation, where \(\gamma\) is the update coefficient:
\begin{equation}
\label{eq:priority}
Priority(i) = (1-\gamma) \cdot Priority(i) + \gamma \cdot score(C')\, , \,\,\,\,\text{if $i \in T_m$}
\end{equation}

\subsection{Heuristic-Based Test Case Generation}
\label{sec:heuristic}
The generated witness test cases should meet both similarity and diversity criteria, which must have execution traces similar to bug-triggering test cases while maintaining diversity among themselves.
Wit-HW generates new test cases by selecting a seed test case and choosing several positions within that test case to mutate in each iteration. Heuristics guide this test generation process.

First, we use the similarity criterion as heuristic guidance for selecting the seed test case for mutation. The fitness of a seed test case \(C\) is defined by its similarity to the bug-triggering test case \(C_b\):
\begin{equation}
Fitness(C) = \frac{1}{1+Distance(C,C_b)}
\end{equation}
Suppose there are \(m\) seed test cases in the seed test case set. The selection probability \(p_s(C)\) for seed test case \(C\) is calculated by:
\begin{equation}
\label{eq:sel-case}
p_s(C) = \frac{Fitness(C)}{\sum_{j=1}^m Fitness(C_j)}
\end{equation}

\begin{figure}[t]
\vspace{-5pt}
\begin{algorithm}[H]
\caption{Heuristic-Based Witness Test Case Generation}
\label{alg:AOS}
\begin{algorithmic}[1] 
    \setlength{\baselineskip}{0.95\baselineskip} 
    \vspace{-3pt}
    \State \textbf{Input:} \(C_b\): Bug-triggering test case 
    \State \textbf{Output:} $\mathcal{P}$: A set of witness test cases
    \State \textbf{Hyper-parameters:} $\text{max\_iterations}, N\color{lightgray}, \alpha, \beta, \gamma, \delta$ (Sec.~\ref{sec:config}) \\
    $\mathcal{S} \gets \{C_b\} $    \quad\quad\quad\quad \cmd{/* Seed test case set */}  \\
    $\mathcal{S'} \gets \{\}$       \quad\quad\quad\quad\,\,\, \cmd{/* Failing test case set */}  \\
    $\mathcal{A}  \gets \{\delta \mid i \in T\}$    \quad\,\cmd{/* Position priority vector */}\\
    $\text{\textit{iter}} \gets 0$  
    \While{\(\text{\textit{iter}}<\text{max\_iterations}\)}
        \State\label{algo:sel-seed-case} $C \gets \text{\textit{select\_seed}}(\mathcal{S})$
        \State\label{algo:sel-mut-pos} $T_m \gets \text{\textit{select\_position}}(C, \mathcal{A})$
        \State\label{algo:do-mutate} $C' \gets \text{\textit{mutate}}(C, T_m )$
        \If {$C'$ is passing $\wedge \, \forall C_i \in \mathcal{S}, Distance(C', C_i) \ne 0$}
            \State $\mathcal{S} \gets \mathcal{S} \cup \{C'\}$
        \ElsIf {$C'$ is failing}
            \State $\mathcal{S'} \gets \mathcal{S'} \cup \{C'\}$
        \EndIf
        \State\label{algo:update-pri} $ \mathcal{A}  \gets \text{\textit{update\_priority}}(C', T_m, \mathcal{S})$
        \State $\text{\textit{iter}} \gets \text{\textit{iter}} + 1$
    \EndWhile
    \If {$\text{Size}(\mathcal{S}) > 0$}
        \State $\mathcal{P} \gets \text{\textit{top\_witness}}(\mathcal{S}, N)$
        \State \text{return} $\mathcal{P}$
    \Else
        \State $\mathcal{S} \gets \mathcal{S'}$
        \State \text{Repeat from Line 5}
    \EndIf
\end{algorithmic}
\end{algorithm}
\vspace{-22pt}
\end{figure}

Second, we apply the diversity criterion as heuristic for selecting positions for mutation. Suppose the test case has $t$ clock cycles, the selection probability \(p_m(i)\) for mutation position \(i\) is calculated by:
\begin{equation}
\label{eq:mut-pos-prob}
p_m(i) = \frac{Priority(i)}{\sum_{j=1}^t Priority(j)}
\end{equation}

Then, we apply mutation operations to the selected time positions within the chosen seed test case. We approach the test case at signal level, similar to some mutation-based hardware test generation techniques~\cite{rfuzz,directfuzz}.
Suppose the design has \( n \) input signals. Test case \( C \) provided to the design contains the values of these \( n \) input signals over \( t \) clock cycles. We can express this as 
\(C=\{(I_1^1,I_2^1,\ldots I_n^1), (I_1^2,I_2^2,\ldots I_n^2), \ldots (I_1^t,I_2^t,\ldots I_n^t)\}\).
When mutating the test case, we select specific time positions and randomly modify several signal values in that clock cycle. Each selected signal value has a probability \( p \) of mutating to a random value within the signal's value range. Let \( T_m \) be the set of clock cycles chosen for mutation. The mutation operation is represented as:
\begin{equation}
\label{eq:mutate}
   I_i^j = 
    \begin{cases} 
    \text{random value}, & \text{if } j \in T_m \text{ with probability } p \\
        I_i^j, & \text{if } j \in T_m \text{ with probability } 1{-}p 
    \end{cases}
\end{equation}



We present the overall process of witness test case generation in Algorithm~\ref{alg:AOS}.
The initial set of seed test cases contains only the bug-triggering test case \(C_b\), and each mutation position starts with an initial priority value \(\delta\). 
Lines 8-19 construct a set of witness test cases until achieving maximum iterations. 
Line~\ref{algo:sel-seed-case} selects a seed test case \(C\) using the probability provided by the similarity heuristic in Eq.~\eqref{eq:sel-case}.
Line~\ref{algo:sel-mut-pos} gets the mutation positions \(T_m\) with the probability provided by the diversity heuristic in Eq.~\eqref{eq:mut-pos-prob}.
Line~\ref{algo:do-mutate} uses \(T_m\) to mutate test case \(C\), generating a new test case \(C'\).
Lines 12-16 evaluate whether \(C'\) is accepted by seed set \(\mathcal{S}\) or failing set \(\mathcal{S}'\) based on its execution results and similarity to existing witness cases in \(\mathcal{S}\).
Lines~\ref{algo:update-pri} update the priority of mutation positions.
Lines 20-26 determine whether stopping the construction process. If witness test cases are generated, the top \(N\) witness test cases with the highest fitness in seed set \(S\) are selected as output. If no witness cases are constructed, the process repeats using generated failing cases as seed test cases.

\subsection{Suspiciousness Calculation}
\label{sec:sbfl}
After constructing a set of witness test cases \(\mathcal{P}\), Wit-HW localizes hardware bugs by analyzing the set of witness test cases and the bug-triggering test case. Following SBFL, Wit-HW computes the suspicious value for each statement within the touched code when executing the given bug-triggering test case. Here Wit-HW adopts Ochiai~\cite{ochiai} as in Eq.~\eqref{eq:ochiai}, one of the most effective formula in SBFL, to compute the suspicious score for each touched statement. 
Here, we only use the given failing test case and just consider the statements touched by the failing test case, and thus \(e_f(s)\) is 1 and \(n_f(s)\) is 0. Therefore Eq.~\eqref{eq:ochiai} can be further simplified into 
$\text{\textit{sus}}(s) = {1}/{\sqrt{1 + e_{p}(s)}}$
in Wit-HW.




\section{Experiments and Results}


\subsection{Benchmarks}
For the evaluation of Wit-HW, we develop the benchmarks based on Tarsel~\cite{tarsel} and RTL-Repair~\cite{rtl_repair}, which consists of 10 designs and 41 bugs, as detailed in Table~\ref{tab:benchmark}. Each bug is paired a bug-triggering test case written by verification engineers.
We categorize the benchmarks into three levels of difficulty. The easy level includes small combinational designs. The medium level features small sequential designs. The hard level contains some industrial designs. Figure~\ref{fig:bug_example} shows the examples of bugs used in our benchmarks. Besides, we also evaluate Wit-HW on 13 real-world bugs in open-source FPGA projects~\cite{fpga_debug_benchmark}, which are detailed in Section~\ref{sec:fpga-bug-exp}.  


\renewcommand{\thetable}{I}
\begin{table}[H]
    \centering
    \renewcommand{\arraystretch}{0.95}
    \vspace{-5pt}
    \caption{Hardware bug localization benchmarks.}
    \vspace{-5pt}
    \label{tab:benchmark}
    \begin{tabular}{c|l|c|c}
        \toprule
        \textbf{Category} & \textbf{Design} & \textbf{Size (LOC)} & \textbf{\#Bugs} \\ \hline
        \multirow{2}{*}{Easy} & decoder\_3\_to\_8 & 25 & 6 \\ \cline{2-4}
                            &  alu  & 37 & 6 \\ \hline
                            
        \multirow{4}{*}{Medium} & counter & 56 & 3 \\ \cline{2-4}
                            &  led\_controller  & 76 & 4 \\\cline{2-4}
                            &  arbiter & 112 & 3 \\ \cline{2-4}
                            &  fsm\_16 & 132 & 4 \\ \hline
                            
        \multirow{4}{*}{Hard} & sdram\_controller & 420 & 3 \\ \cline{2-4}
                            &  sha3 & 499 & 3 \\ \cline{2-4}
                            &   i2c  & 2018 & 6 \\ \cline{2-4}
                            &  reed\_decoder & 4366 & 3 \\
        \bottomrule
    \end{tabular}
\end{table}

\vspace{-30pt}

\lstset{
    columns = flexible,       
    basicstyle = \linespread{1}
    \fontfamily{ascii}\selectfont\fontsize{5.8}{6.5}\selectfont,
    numbers = none,                                      
    numberstyle = \tiny \color{gray},                    
    keywordstyle = \bfseries, 
    frame=leftline,
    showstringspaces = false,                            
    language=Verilog                                         
}
\begin{figure}[h]
    \centering
    \begin{multicols}{3}
        \begin{subfigure}[t]{0.14\textwidth}
            \centering
            \begin{lstlisting}
always@(*)
  case(opcode)
    // buggy
    4'b0000: y = a-b;
    4'b0001: y = a+b;
    // correct
    4'b0000: y = a+b;
    4'b0001: y = a-b;
            \end{lstlisting}
            \vspace{-3pt}
            \caption{ALU Bug-1 (Easy)}
        \end{subfigure}
        
        \begin{subfigure}[t]{0.15\textwidth}
            \centering
            \begin{lstlisting}
always@(posedge clk)
  ...
  if(state==S11)
    if(in1&in2)
      // buggy
      state <= S6;
      // correct
      state <= S7;
            \end{lstlisting}
            \vspace{-3pt}
            \caption{FSM Bug-2 (Med.)}
        \end{subfigure}

        \begin{subfigure}[t]{0.14\textwidth}
            \centering
            \begin{lstlisting}
always@(posedge clk)
  ...
  case(state)
    start_b:
      // buggy
      scl_oen <= 1'b0;
      // correct
      scl_oen <= 1'b1;
            \end{lstlisting}
            \vspace{-3pt}
            \caption{I2C Bug-5 (Hard)}
        \end{subfigure}
    \end{multicols}
    \vspace{-10pt}
    \caption{Example of hardware design bugs in benchmarks.}
    \vspace{-15pt}
    \label{fig:bug_example}
\end{figure}

\subsection{Experimental Setup}
\subsubsection{Implementation} Wit-HW uses Verilator~\cite{verilator} v5.027 as simulator to observe design behaviors under given test cases. We generate line coverage reports and VCD wave files using Verilator to calculate the coverage and state distance. 
In our experiments, we employ the correct RTL design as the output validator. Specifically, the validator compares the outputs of the correct and buggy designs cycle by cycle. If their outputs diverge, a bug is deemed triggered. When localizing bugs, we rely only on the output behaviors of the correct design as a reference for correctness, without accessing any internal information.

\subsubsection{Configuration}\label{sec:config} 
In Eq.~\eqref{eq:distance}, we balance the weight of coverage distance and state distance by setting \(\alpha\) to \(1/z\) and set \(\beta\) to \(1/k\), where \(z\) is the number of coverpoints and \(k\) is the number of registers.
We define the mutation probability \(p\) in Eq.~\eqref{eq:mutate} as \(0.5\), and set the priority update coefficient \(\gamma\) in Eq.~\eqref{eq:priority} as \(0.1\), with the initial mutation priority \(\delta\) set to \(1.0\). 
In Algorithm~\ref{alg:AOS}, the maximum test generation iterations are set to \(100\), and the top \(10\) witness test cases with the highest fitness values are selected for suspiciousness calculation.
Our study is conducted on a workstation with 754 GiB of RAM and two Intel Xeon Gold 6248R CPUs.

\subsubsection{Measurement}
To evaluate the effectiveness of bug localization, we use two metrics: Top-n and Mean Average Rank (MAR). 
Top-n measures the number of successfully localized bugs within the Top-n positions (\ie \(n \in \{1,5,10,20\}\) in our study). Higher Top-n values indicate better performance.
MAR calculates the average localization rank of all the bugs. Lower MAR values indicate more accurate localization. The lower bound of MAR value is \(1\), which means all bugs are localized at Top-1.
If multiple statements have the same suspiciousness, we assign them their average rank. For designs with multiple bugs, we consider the highest rank among them.

\subsubsection{Compared Baseline}
We compare Wit-HW with the bug localization framework Tarsel~\cite{tarsel} and the bug repair framework RTL-Repair~\cite{rtl_repair}.
When reproducing Tarsel, we sample the coverage data before and after the initial bug trigger for suspicion analysis. 
For RTL-Repair, since it directly repairs the design and lacks precise localization techniques to calculate ranks, we measure successful bug repairs as Top-1. If RTL-Repair fails to repair the bug, we assign the rank equal to half the number of suspicious statements from the repair area, consistent with prior work~\cite{tarsel}.
Additionally, to assess the impact of our heuristic-based test generation strategy, we replace the search strategy with a random approach and do not use guidance for constructing witness test cases. We call this variant as Wit-HW\(_{rand}\).

\renewcommand{\thetable}{II}
\begin{table*}[t]
    \renewcommand{\arraystretch}{0.75}
    \centering
    \caption{Hardware bug localization effectiveness comparison.}
    \vspace{-5pt}
    \begin{threeparttable}
    \begin{tabular}{c|c|cc|c c|c c|c c|c c}
    \toprule 
    \textbf{Category} & \textbf{Approach} & \textbf{Top-1} &\textbf{$\%_\textnormal{Top-1}$} & \textbf{Top-5}& \textbf{$\%_\textnormal{Top-5}$} & \textbf{Top-10}& \textbf{$\%_\textnormal{Top-10}$} & \textbf{Top-20} & \textbf{$\%_\textnormal{Top-20}$} &\textbf{MAR} & \textbf{$\Uparrow_\mathit{MAR}$}  \\

    \midrule 
    \multirow{4}{*}{ \textbf{Easy} }
    & Wit-HW                       & 9  & 75\%  & 11 & 92\% & 12 & 100\% & 12 & 100\% & 1.75 & 1.00$\times$ \\
    & Tarsel                       & 9  & 75\% & 11 & 92\% & 12 & 100\% & 12 & 100\% & 1.83 & 1.05$\times$ \\
    & RTL-Repair                   & 3  & 25\% & 3  & 25\% & 6 & 50\% & 12 & 100\%  & 8.62 & 4.93$\times$ \\
    & Wit-HW\(_{rand}\)            & 8  & 67\% & 11 & 92\% & 12 & 100\% & 12 & 100\%  & 2.41 & 1.38$\times$ \\
    
    \midrule 
    \multirow{4}{*}{ \textbf{Medium} }
    & Wit-HW                        & 8  & 57\% & 11 & 79\% & 14 & 100\% & 14 & 100\% & 2.78 & 1.00$\times$ \\
    & Tarsel                        & 0  & \ 0\% & 3 & 21\%   & 6 & 43\% & 10 & 71\%  & 14.92 & 5.37$\times$ \\
    & RTL-Repair                    &  2 & 14\% & 3 & 21\%  & 6 & 43\% & 10 & 71\% & 13.64 & 4.91$\times$ \\
    & Wit-HW\(_{rand}\)             & 6  & 43\% & 9 & 64\%  & 12 & 86\% & 14 &  100\% & 5.35 & 1.92$\times$ \\
    \midrule 
    \multirow{4}{*}{ \textbf{Hard} }
    & Wit-HW                        & 3  & 20\%  & 8 & 53\% & 10 & 67\% & 10 & 67\%  & 42.13 & 1.00$\times$ \\
    & Tarsel                        & 0  & \ 0\%   & 2 & 13\% & 4  & 27\% & 4 & 27\%  & 81.00 & 1.92$\times$ \\
    & RTL-Repair                    & 1  & \ 7\%   & 1 & \ 7\%  & 4  & 27\% & 4 & 27\%  & 73.93 & 1.75$\times$ \\
    & Wit-HW\(_{rand}\)             & 2  & 13\%  & 5 & 33\% & 8  & 53\% & 9 & 60\%  & 49.93 & 1.19$\times$ \\
    \midrule 
    \multirow{4}{*}{ \textbf{Overall} }
    & Wit-HW                        & 20  & 49\%  & 30 & 73\% & 36 & 88\% & 36 & 88\% & 16.90  & 1.00$\times$ \\
    & Tarsel                       & 9  & 22\%    & 16 & 39\% & 22 & 54\% & 26 & 63\% & 35.26 & 2.09$\times$ \\
    & RTL-Repair                   & 6  & 15\%    & 7  & 17\% & 16 & 39\% & 26 & 63\% & 34.23 & 2.03$\times$ \\
    & Wit-HW\(_{rand}\)             & 14  & 34\%  & 24 & 59\% & 33 & 80\% & 36 & 88\%  & 20.80 & 1.23$\times$ \\
    
    \bottomrule
    \end{tabular}
  \end{threeparttable}
  \label{tab:main-exp}
  \vspace{-10pt}
\end{table*}

\subsection{Results and Analysis}
\label{sec:exp_1}
\subsubsection{Overall Effectiveness of Wit-HW} 
Table~\ref{tab:main-exp} presents the effectiveness of Wit-HW. Overall Wit-HW successfully localize 20\,/\,30\,/\,36 bugs (out of 41 bugs) within Top-1\,/\,5\,/\,10 ranks, which substantially outperforms Tarsel~\cite{tarsel} (9\,/\,16\,/\,22) and RTL-Repair~\cite{rtl_repair} (6\,/\,7\,/\,16), demonstrating the effectiveness of introducing witness test cases for hardware bug localization.
We further analyze the effectiveness of Wit-HW for different benchmark subjects. For easy subject which consists of combinational designs, both Wit-HW and Tarsel perform well. However, for medium complexity subject which involves sequential designs, the performance of Tarsel degrades significantly. 
RTL-Repair also exhibits poor performance in both easy and medium subjects due to its lack of a precise localization process. 
Moreover, we find that Wit-HW can localize 8 bugs (out of 12 bugs) in Top-5 for hard designs, significantly outperforming Tarsel and RTL-Repair, which demonstrates the scalability of Wit-HW. The MAR for hard designs is higher than for medium and easy benchmarks. That is because in several bug cases, all the methods cannot accurately localize the bug, resulting in extreme outlier ranks due to large number of statements in hard designs, which inflates the average rank.
When comparing with Wit-HW\(_{rand}\), Wit-HW outperforms Wit-HW\(_{rand}\) in terms of all subjects, which indicates that our heuristic-based search strategy is more effective than a random search approach.

\subsubsection{Analysis of Component Effects}
Wit-HW uses coverage distance and state distance as similarity metrics and implements a diversity-driven mutation method. 
To assess the impact of each component, we respectively remove them from Wit-HW. Table~\ref{tab:ablation} presents the experimental results on the overall benchmarks, comparing the original Wit-HW with its ablated versions: Wit-HW without coverage distance (Wit-HW w/o Cov) by setting $\alpha{=}0$ in Eq.~\eqref{eq:distance}, Wit-HW without state distance (Wit-HW w/o State) by setting $\beta{=}0$ in Eq.~\eqref{eq:distance}, and Wit-HW without diversity guidance (Wit-HW w/o Div) by setting $p_m{=}1{/}t$ in Eq.~\eqref{eq:mut-pos-prob}.
The results demonstrate that each component contributes to the overall performance.
We find that the version without coverage distance shows the greatest performance decline. This suggests that coverage reflects most, but not all, hardware design behaviors in the process of bug localization. Meanwhile, state distance and diversity heuristics further enhance performance of Wit-HW by incorporating sequential behavior analysis and reducing data bias.

\renewcommand{\thetable}{III}
\begin{table}[ht]
\vspace{-5pt}
    \centering
    \renewcommand{\arraystretch}{0.9}
    \caption{Effects of removing components from Wit-HW.}
    \vspace{-5pt}
    \label{tab:exp-distance}
    \begin{tabular}{l|ccc|cc@{}}
        \toprule
        \textbf{Approach} & \textbf{Top-1} & \textbf{Top-5} & \textbf{Top-10} & \textbf{MAR} & $\Uparrow_{MAR}$ \\ \midrule
        Wit-HW & 20 & 30 & 36 & 16.90 & 1.00\(\times\) \\
        Wit-HW w/o Cov & 12 & 24 & 32 & 20.41 & 1.21\(\times\) \\
        Wit-HW w/o State & 17 & 24 & 32 & 19.48 & 1.15\(\times\) \\
        Wit-HW w/o Div & 15 & 28 & 34 & 18.02 & 1.07\(\times\) \\ \bottomrule
    \end{tabular}
    \vspace{-5pt}
    \label{tab:ablation}
\end{table}

\begin{figure}[b]
    \vspace{-15pt}
    \centering
    \begin{subfigure}[b]{0.155\textwidth}
        \includegraphics[width=\textwidth]{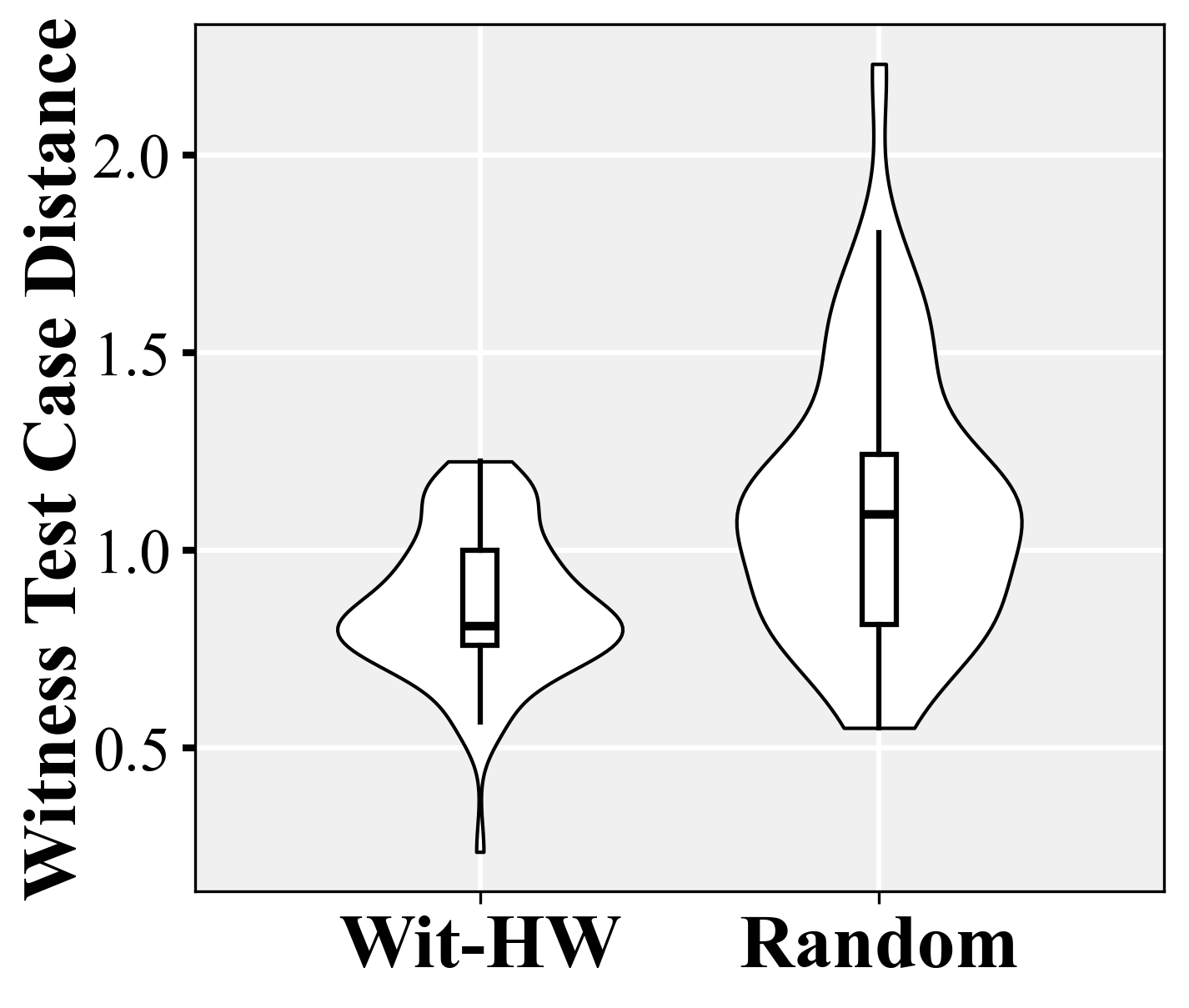}
        \caption{Easy}
        \label{fig:image1}
    \end{subfigure}
    \hfill
    \begin{subfigure}[b]{0.15\textwidth}
        \includegraphics[width=\textwidth]{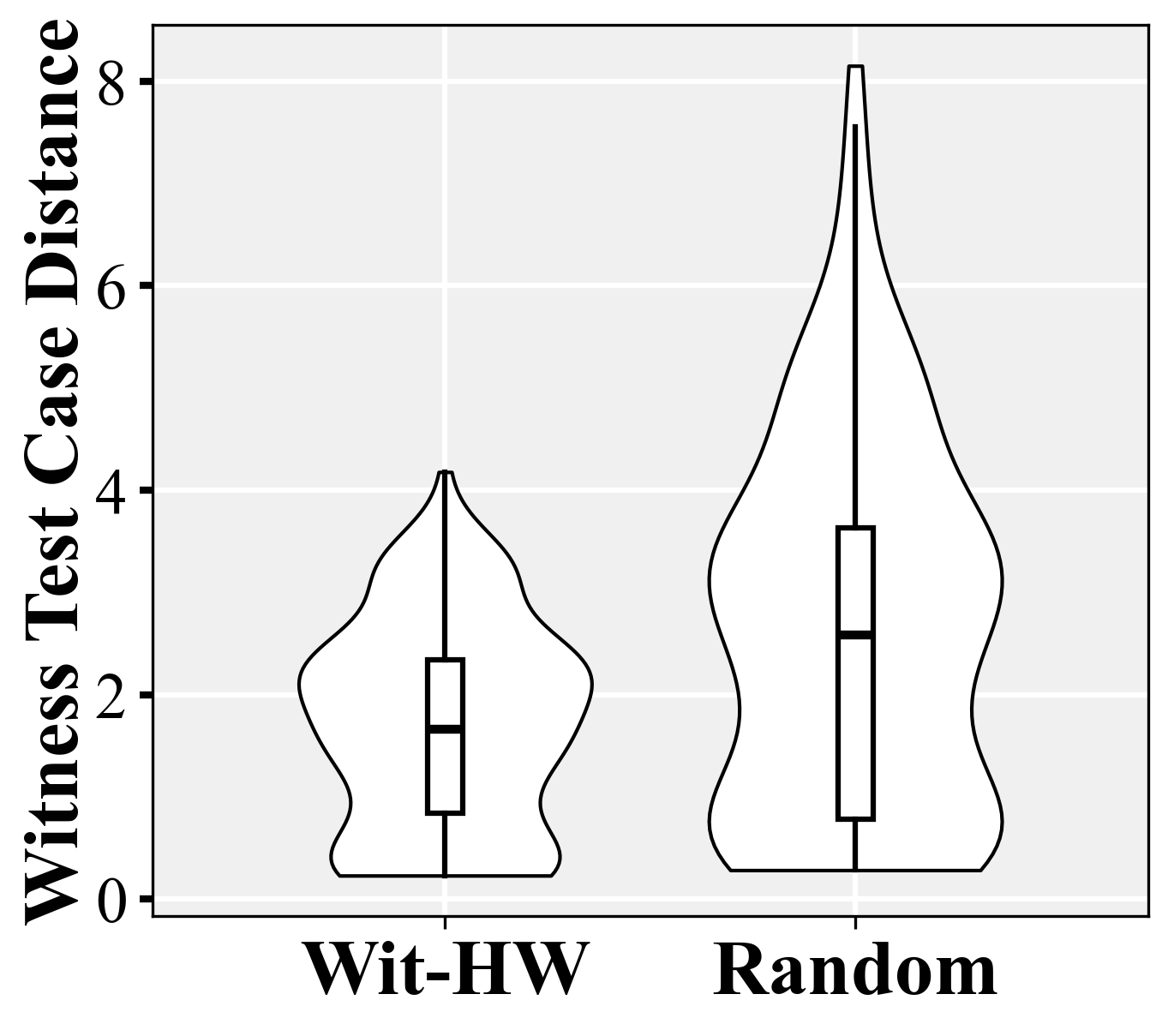}
        \caption{Medium}
        \label{fig:image2}
    \end{subfigure}
    \hfill
    \begin{subfigure}[b]{0.155\textwidth}
        \includegraphics[width=\textwidth]{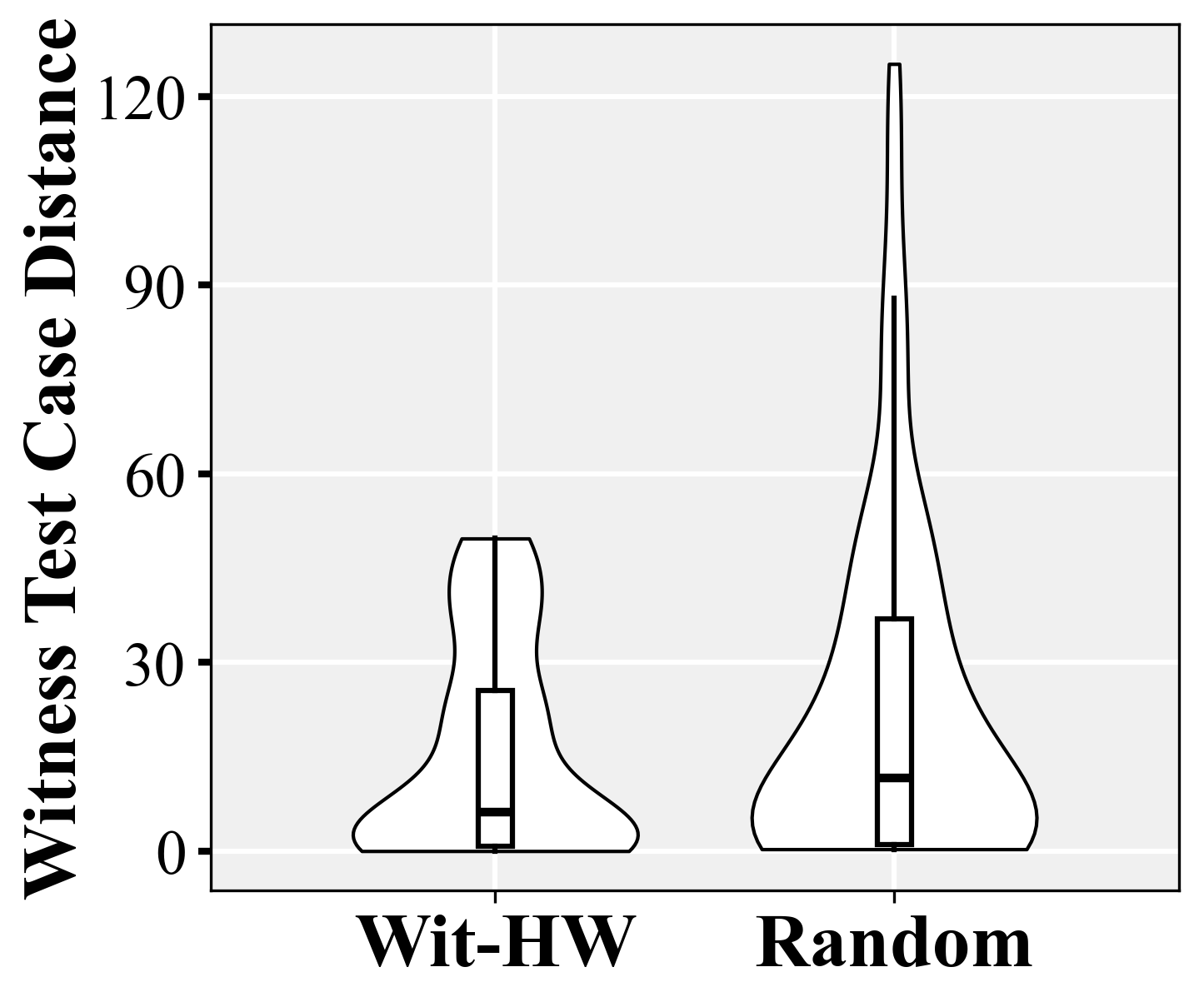}
        \caption{Hard}
        \label{fig:image3}
    \end{subfigure}
    \vspace{-5pt}
    \caption{Distance between generated witness test cases and the initial bug-triggering test case using Wit-HW and Wit-HW\(_{rand}\).}
    \label{fig:distance_exp}
\end{figure}

\subsubsection{Analysis of Distance Metric}
Wit-HW assumes that witness test cases exhibiting similar hardware execution behavior to the bug-triggering test case can help exclude more innocent code statements and improve bug localization accuracy. 
To explore why witness test cases generated by Wit-HW achieve better bug localization performance in SBFL, we conduct a quantitative evaluation of the similarity between the generated witness test cases and the bug-triggering test case. 
The similarity is measured using the distance metric defined in Eq.~\eqref{eq:distance}.
Figure~\ref{fig:distance_exp} illustrates the distance to the bug-triggering test case of the witness test cases generated by Wit-HW and Wit-HW\(_{rand}\).
The violin plots depict the density of distances across different values, and the box plots display the median and interquartile ranges. 
The results indicate that Wit-HW produces witness test cases with greater similarity to the bug-triggering test case, thereby enhancing the precision of bug localization. The heuristic-driven approach of Wit-HW ensures that the generated test cases align more closely with the execution behavior of the faulty scenario, leading to improved fault isolation and reduced noise in the localization process.

\subsubsection{Analysis of Iteration Number} 
\begin{figure}[h]
    \centering
    \vspace{-5pt}
    \begin{subfigure}{0.23\textwidth}
        \vspace{-2pt}
        \includegraphics[width=\textwidth]{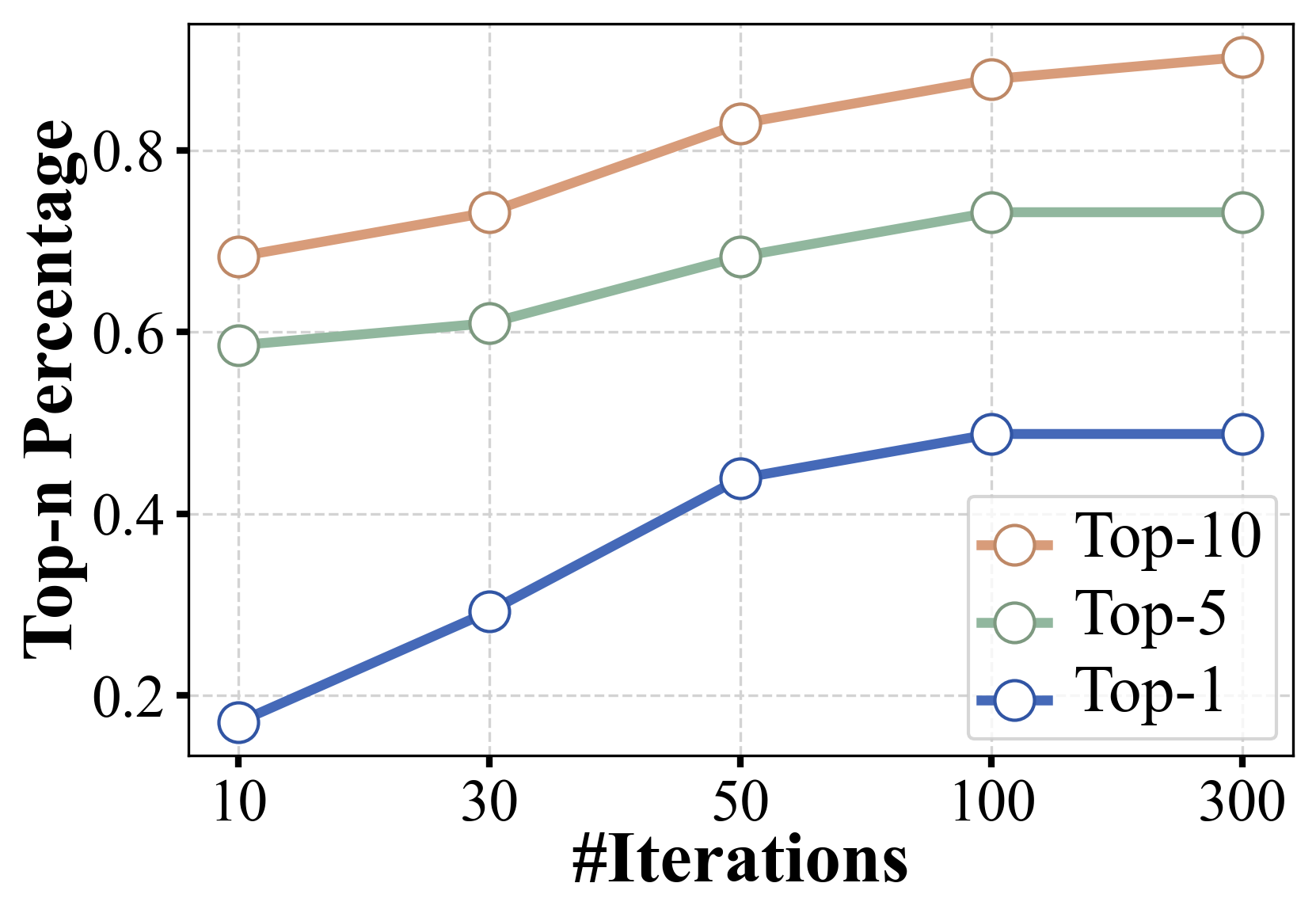}
    \end{subfigure}
    \begin{subfigure}{0.23\textwidth}
        \includegraphics[width=\textwidth]{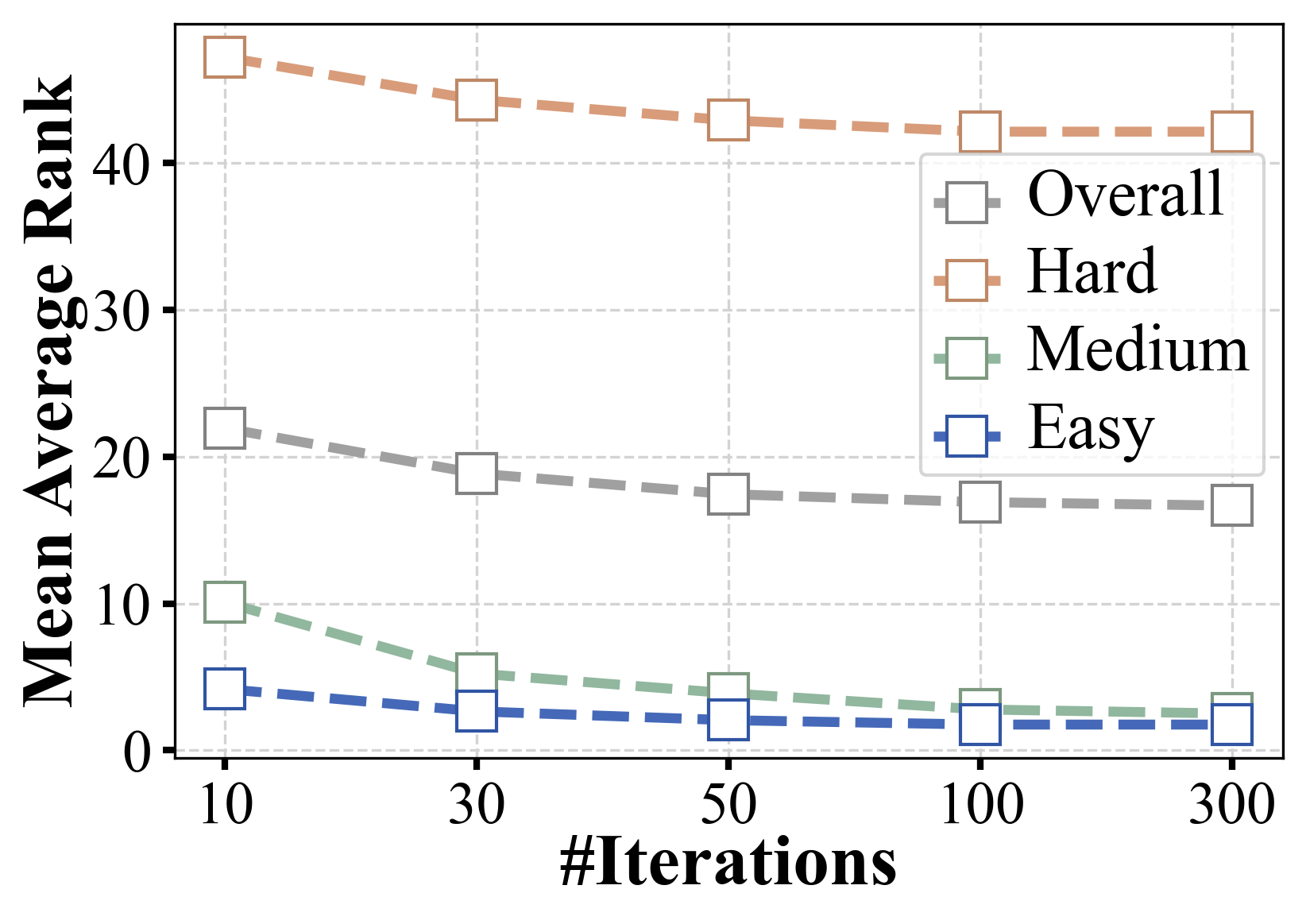}
    \end{subfigure}
    \vspace{-5pt}
    \caption{Effects of iteration times for Wit-HW.}
    \vspace{-5pt}
    \label{fig:iter-exp}
\end{figure}

Wit-HW uses a heuristic-based test generation approach, which relies on iterative simulations to progressively generate effective witness test cases. We evaluate the impact of iteration times on the performance of Wit-HW. Figure~\ref{fig:iter-exp} presents the experimental results. The left sub-figure shows the Top-n bug localization percentage of Wit-HW on overall benchmarks. The right sub-figure shows the Mean Average Rank (MAR) of Wit-HW for each difficulty level in the benchmarks.  
During the initial phases of test case generation, incremental increases in iterations lead to substantial improvements in bug localization accuracy. However, when the number of iterations reaches approximately \(100\), further increases do not significantly enhance bug localization performance. This observation explains our choice of \(100\) iterations in the experiments, as it achieves the upper performance limit of Wit-HW while minimizing the number of simulation iterations.


\subsubsection{Speed Comparison}
Table~\ref{tab:exp-speed} presents the time taken and the success rate of Top-1 bug localization for Wit-HW and its comparison methods. Since different testbenches are required to trigger different bugs within the same design, the table reports the average testbench cycle length and the average time cost for each method across all designs.
For Wit-HW, simulation dominates the runtime, which is influenced by the length of testbench cycles, as well as the scale and complexity of the hardware design. 
Wit-HW employs an iterative test generation process and requires simulations for each newly generated test case. This iterative approach results in longer runtime compared to Tarsel~\cite{tarsel} and RTL-Repair~\cite{rtl_repair}. 
However, the overhead is justified, as it significantly enhances the accuracy of bug localization. By refining effective witness test cases iteratively, Wit-HW achieves a higher success rate in identifying hardware bugs, which reduces the overall debugging effort and accelerates the development process. 
\renewcommand{\thetable}{V}
\begin{table}[h]
\centering
\renewcommand{\arraystretch}{0.85}
\caption{Time and success rate of Top-1 bug localization for Wit-HW and comparison methods. ``\#Cy.'' shows the number of clock cycle steps in the provided testbench.}
\vspace{-5pt}
\label{tab:exp-speed}
\begin{tabular}{p{1.6cm}>{\centering\arraybackslash}p{0.8cm}|>{\centering\arraybackslash}p{0.5cm}>{\centering\arraybackslash}p{0.7cm}|>{\centering\arraybackslash}p{0.4cm}>{\centering\arraybackslash}p{0.5cm}|>{\centering\arraybackslash}p{0.4cm}>{\centering\arraybackslash}p{0.4cm}}
\toprule
\textbf{Design} & \textbf{\#Cy.} & \multicolumn{2}{c}{\textbf{Wit-HW}} & \multicolumn{2}{|c}{\textbf{Tarsel}} & \multicolumn{2}{|c}{\textbf{RTL-Repair}} \\ 
\midrule
decoder\_3\_to\_8   & 8     &  \textbf{5\,/\,6} & 32.3\,s   & \textbf{5\,/\,6} & 1.5\,s    & 3\,/\,6 & 1.0\,s \\
alu                 & 7     &  \textbf{4\,/\,6} & 32.9\,s   & \textbf{4\,/\,6} & 1.3\,s    & 0\,/\,6 & 1.4\,s \\
counter             & 9     &  \textbf{2\,/\,3} & 35.0\,s   & 0\,/\,3 & 1.1\,s    & \textbf{2\,/\,3} & 1.1\,s \\
led\_controller     & 10    &  \textbf{3\,/\,4} & 35.5\,s   & 0\,/\,4 & 1.5\,s    & 0\,/\,4 & 1.6\,s \\
arbiter             & 8     &  0\,/\,3 & 37.3\,s   & 0\,/\,3 & 1.4\,s    & 0\,/\,3 & 6.2\,s \\
fsm\_16             & 6     &  \textbf{3\,/\,4} & 33.6\,s   & 0\,/\,4 & 0.9\,s    & 0\,/\,4 & 0.8\,s \\
sdram\_controller   & 58    &  \textbf{1\,/\,3} & 59.9\,s   & 0\,/\,3 & 1.3\,s    & 0\,/\,3 & 2.6\,s \\
sha3                & 46    &  0\,/\,3 & 54.5\,s   & 0\,/\,3 & 2.6\,s    & 0\,/\,3 & 1.9\,s \\
i2c                 & 719   &  \textbf{1\,/\,6} & 49.5\,s   & 0\,/\,6 & 1.9\,s    & 0\,/\,6 & 2.0\,s \\
reed\_decoder       & 1566  &  \textbf{1\,/\,3} & 86.7\,s   & 0\,/\,3 & 3.0\,s    & \textbf{1\,/\,3} & 2.5\,s \\
\midrule
Total \& Avg. & $-$ & \textbf{20\,/\,41} & 45.7\,s & 9\,/\,41 & 1.7\,s & 6\,/\,41 & 2.1\,s \\
\bottomrule
\end{tabular}
\vspace{-15pt}
\end{table}

\subsection{Open-Source Bug Localization}
\label{sec:fpga-bug-exp}
In addition to the evaluation benchmarks used in Tarsel~\cite{tarsel} and RTL-Repair~\cite{rtl_repair}, which are specifically created to assess automatic bug localization and repair tools, we also apply Wit-HW to a set of real-world bugs extracted from git commits in open-source FPGA hardware projects~\cite{fpga_debug_benchmark}. 
We use 13 out of the 21 reproducible bugs provided by previous work~\cite{fpga_debug_benchmark}. The other 8 cases contain SystemVerilog features that are unsupported by our simulator, or lack a definitive ground-truth bug position.

Table~\ref{tab:exp-fpga} presents the details of bug cases and the localization results, including description of the bugs, project sizes (in lines of code), sizes of the buggy sections (in lines of code), and the length of provided testbench (in clock cycles).
We evaluate the quality of bug localization based on the following scale: (A) localized at rank 1, (B) tied for rank 1, (C) localized in the Top-10, and (D) unable to accurately localize the bug.

\renewcommand{\thetable}{VI}
\begin{table}[h]
\centering
\renewcommand{\arraystretch}{0.85}
\caption{Wit-HW localization results for bugs in open-source FPGA project. ``LOC'' indicates the lines of code in the project. ``Diff" indicates the number of lines that differ from the ground truth in the buggy design. ``\#Cy.'' shows the number of clock cycle steps in the provided testbench.}
\label{tab:exp-fpga}
\begin{tabular}{>{\arraybackslash}p{0.4cm}c|>{\centering\arraybackslash}p{0.4cm}>{\centering\arraybackslash}p{0.4cm}>{\centering\arraybackslash}p{0.45cm}|>{\arraybackslash}p{0.45cm}c>{\arraybackslash}p{0.4cm}c}
\toprule
\textbf{Project} & \textbf{Bug Description} & \textbf{LOC} & \textbf{Diff} & \textbf{\#Cy.} & \textbf{Time} & \textbf{Quality} \\
\midrule
D13 & Failure to update & 122 & 3 & 6 & 37.3\,s & A \\
C1 & Dead lock & 1284 & 1 & 35k & 303.5\,s & B \\
C4 & Signal asynchrony & 547 & 1 & 10 & 40.6\,s & B \\
D9 & Endianness mismatch & 1284 & 2 & 924 & 92.2\,s & B \\
D11 & Failure to update & 146 & 2 & 17 & 25.7\,s & B \\
S2 & Protocol violation & 232 & 1 & 45 & 38.8\,s & B \\
\midrule
D12 & Failure to update & 341 & 1 & 16 & 43.1\,s & C \\
S1.R & Protocol violation & 731 & 1 & 10 & 37.8\,s & C \\
S1.B & Protocol violation & 731 & 2 & 10 & 40.9\,s & C \\
\midrule
C3 & Data control asynchrony & 1284 & 7 & 501 & 43.8\,s & D \\
D4 & Buffer overflow & 404 & 27 & 185 & 172.3\,s & D \\
D8 & Misindexing & 729 & 2 & 14 & 29.2\,s & D \\
S3 & Incomplete implement & 475 & 35 & 13 & 29.0\,s & D \\
\bottomrule
\end{tabular}
\vspace{-5pt}
\end{table}

The experiment results show that Wit-HW localizes 6 out of 13 bugs at rank 1 or tied rank 1. Wit-HW also successfully localizes 3 bugs within the Top-10 ranks. 
Compared to the experiment results presented in RTL-Repair~\cite{rtl_repair}, which can only correctly fix 2 bugs in the same benchmarks, our bug localization method achieves a broader applicability. 
Although Wit-HW focuses on bug localization and does not consider the automatic repair like RTL-Repair~\cite{rtl_repair}, it is sufficient for real-world debugging, where identifying and analyzing the root cause of bugs is the primary challenge~\cite{synopsys_study}.
Additionally, RTL-Repair lacks an exact bug localization process. It directly uses variable-included templates to modify large regions of buggy designs and then employs a formal solver to achieve the repair.
In contrast, Wit-HW precisely localizes hardware bugs to a relatively small region. This shows the potential for integrating Wit-HW bug localization into automatic repair processes, which could significantly narrow down the suspicious repair areas and enhance the repair success rate.

We also find that Wit-HW is effective for designs with fewer bugs. When there are too many bugs in the design (\eg C3, D4, S3), Wit-HW struggles to generate successful witness test cases, or the generated witness test cases may not be effective. These cases often diverge significantly from the original bug-triggering test case, making them unsuitable for spectrum-based fault localization.
In comparison to the experimental results in Section~\ref{sec:exp_1}, the benchmarks from real-world open-source FPGA projects present increased complexity and diverse bug-triggering scenarios, which inevitably reduce the localization accuracy of Wit-HW. We will discuss these limitations in the following section.

\section{Limitation and Discussion}
In our experiment, we set the maximum iterations to \(100\). Beyond this limit, increasing the iterations do not significantly enhance bug localization performance. In this section, we analyze the unresolved bugs and identify the limitations and future directions of Wit-HW.

\begin{figure}[t] 
    \centering 
    \includegraphics[width=0.48\textwidth]{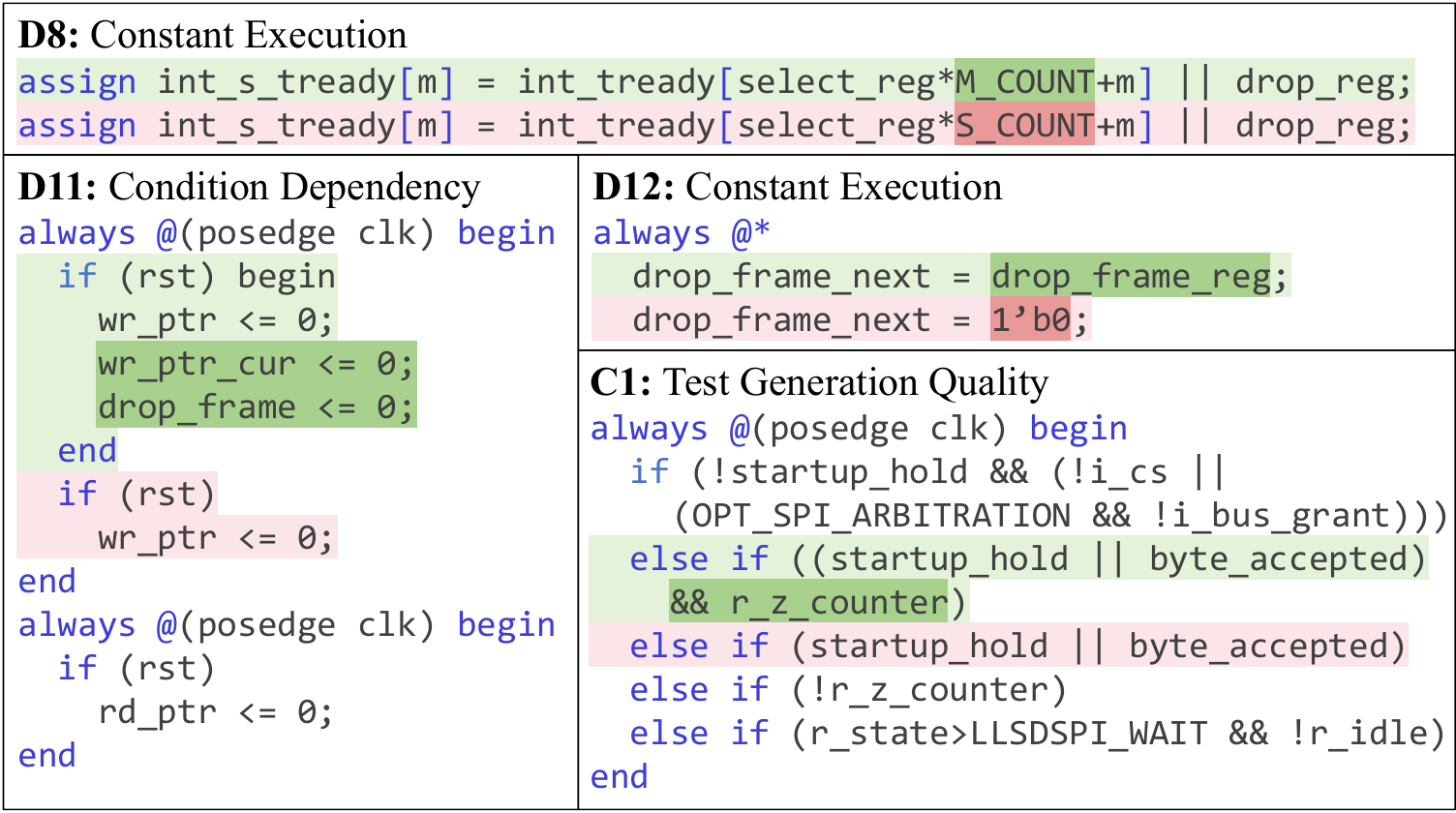} 
    \caption{Examples of buggy cases that cannot be localized by Wit-HW. The red segments indicate the buggy statements, and the green segments indicate the original, correct logic of the buggy statements.}
    \vspace{-10pt}
    \label{fig:bug_case} 
\end{figure}

\textbf{Constant Execution.}
Wit-HW uses spectrum-based method that analyze coverage differences between passing and failing test cases to identify bugs. However, some bugs are located at logical paths that are constantly executed. 
For D8 shown in Figure~\ref{fig:bug_case}, the bug is in the assign statement, which is executed in both passing and failing test cases. Similarly, in D12, the bug is a blocking assignment statement outside any conditional scope within an always block, also executed universally. These types of bugs present challenges for Wit-HW.

\textbf{Condition Dependency.}
In hardware design, the execution conditions of certain statements can sometimes be interdependent. When the execution condition of a buggy statement is equivalent to or a sufficient condition for an innocent statement, spectrum-based methods may struggle to localize the exact bug position.
For example, for D11 shown in Figure~\ref{fig:bug_case}, the bug lies in the reset branch of the first always block, where the reset for \cmd{wr\_ptr\_cur} and \cmd{drop\_frame} is neglected. 
To generate passing witness test cases, one should avoid resetting the design while these variables are active. 
However, this also prevents the reset branch in the second always block from executing.
Due to this condition dependency, both reset branches exhibit the same execution trace in the witness test cases. Consequently, spectrum-based localization methods cannot precisely identify the buggy statement, resulting in tied suspicion scores.

The above two limitations of constant execution and condition dependency stem from the inherent constraints of spectrum-based bug localization technique, which relies on the differences in execution traces between passing and failing test cases to identify bugs. When buggy statements do not exhibit execution differences from other statements, they receive the same suspicious score. 
This accounts for the occurrence of some tied ranks observed in the open-source bug localization in Table~\ref{tab:exp-fpga}.
To address these limitations, integrating with mutation-based bug localization technique~\cite{mutation_bug_loc_1, mutation_bug_loc_2, mutation_bug_loc_3} presents a promising solution. This method analyzes bug by mutating design code and observing behaviors of mutant designs.
The suspicious region identified by Wit-HW could be further used as candidate regions for mutation, allowing for more precise bug identification.

\textbf{Test Generation Quality. }
The performance of Wit-HW is also constrained by the current test generation method. 
We observe that Wit-HW exhibits less accurate bug localization and struggles to generate effective witness test cases for complex buggy designs compared to simpler cases. 
One reason is our use of a signal-level mutation strategy, which can easily produce invalid input patterns for complex designs with strict and intricate interface protocols. 
Additionally, for buggy branches with complex conditions, generating effective witness test cases through design-agnostic test case mutation is challenging. For instance, in the C1 buggy case in Figure~\ref{fig:bug_case}, the bug resides in an if-else block with several complex conditions. Bypassing this buggy block requires satisfying intricate condition combinations, which is difficult to achieve with simple test case mutation.

To address these challenges, several improvements can be considered. 
First, adopting higher-level mutation strategies (e.g., mutating the sequence of input transactions)~\cite{hw_fuzz, difuzz, noc_fuzz} could enhance the ability to generate valid and effective witness test cases for designs with complex input protocols.
Furthermore, advanced hardware test generation methods, such as concolic testing~\cite{concolic_1, concolic_2}, LLM-aided test generation~\cite{llm4dv, verilogreader}, and hardware fuzzing~\cite{rfuzz, directfuzz, fast_fuzz}, could be explored. 
We have demonstrated the feasibility of framing the hardware bug localization problem as a test generation problem. This opens the door to leveraging diverse hardware test generation methods to enhance the bug localization performance in the future. 


\section{Related Work}
Wit-HW is the only tool that localizes hardware bugs by generate diverse witness test cases.
While there are many bug localization approaches for hardware~\cite{tarsel, veribug, mutation_bug_loc_3, hw_bug_loc_pur, hw_bug_loc_con}, they typically rely on a single testbench as the bug-triggering input. This limits comprehensive observation of design behaviors and impacts the performance. 
Some approaches attempt to repair buggy hardware designs directly without precise bug localization~\cite{cirfix, rtl_repair, llm_fix_1, llm_fix_2, strider}, leading to poor repair applicability. Our approach can serve as a preliminary stage, providing suspicious regions for automatic repair.

Recent works use a symbolic approach for Quick Error Detection (QED) in bug localization~\cite{qed_1, qed_2, qed_3}. This method aims to generate the shortest and deterministic error trace from a lengthy bug-triggering test case. 
While it cannot directly pinpoint the exact bug position and still requires human analysis, there is potential for Wit-HW to utilize these concise error traces to generate witness test cases. This could improve mutation effectiveness and speed up simulation iterations, making it a promising area for future work.

\section{Conclusion}
In this paper, we innovatively formulate hardware bug localization as the problem of witness test case generation. We introduce a novel hardware bug localization technique, Wit-HW, which generates a set of witness test cases to assist in localizing hardware bugs. 
We establish criteria for effective witness test cases and design a heuristic-based test generation method to produce these cases. The experimental results show that Wit-HW successfully localize 49\%\,/\,73\%\,/\,88\% bugs within Top-1\,/\,Top-5\,/\,Top-10 ranks in designs scaling from dozens to thousands of lines of code, significantly outperforming state-of-the-art bug localization techniques. Besides, we evaluate Wit-HW on real-world bugs in open-source FPGA projects and showcase its applicability in practical scenarios. This work offers new insights into the problem of hardware bug localization and opens avenues for employing advanced test generation methods to achieve more efficient and scalable hardware debugging solutions.

\section*{Acknowledgement}
This work was supported by the National Key R\&D Program of China (Grant No. 2022YFB4500500).

\clearpage
\bibliographystyle{IEEEtran}
\bibliography{long}

\end{CJK*}
\end{document}